\documentclass[10pt, a4paper]{article}

\usepackage{lrec-coling2024} 

\usepackage{times}
\usepackage{latexsym}
\usepackage[T1]{fontenc}
\usepackage[utf8]{inputenc}
\usepackage{microtype}
\usepackage{inconsolata}

\usepackage{arydshln}
\usepackage{tabularray}
\usepackage{multirow}
\usepackage{booktabs, caption, array}
\usepackage{algorithm}
\usepackage{algorithmic}
\usepackage{comment}
\usepackage{xcolor}
\usepackage{amssymb}
\usepackage{subfigure}
\usepackage{graphicx}
\usepackage{enumitem}
\usepackage{framed}
\usepackage{amsmath}
\definecolor{shadecolor}{gray}{0.9}
\usepackage{comment}
\newcommand{\ywj}[1]{{\color{violet} ywj: ``#1''}}
\renewcommand{\ywj}[1]{#1}
\usepackage{xcolor}
\usepackage{hyperref}
 \definecolor{darkblue}{rgb}{0, 0, 0.5}
  \hypersetup{colorlinks=true, citecolor=darkblue, linkcolor=darkblue, urlcolor=darkblue}

\usepackage{xstring}

\usepackage{color}

\title{Logic Rules as Explanations for Legal Case Retrieval\\ \vspace*{.5\baselineskip}} 
\name{Zhongxiang Sun\textsuperscript{1}, Kepu Zhang\textsuperscript{1}, Weijie Yu\textsuperscript{2}\textsuperscript{*}\thanks{Corresponding author: Weijie Yu.}, Haoyu Wang\textsuperscript{1}, Jun Xu\textsuperscript{1}} 

\address{\textsuperscript{1}Gaoling School of Artificial Intelligence, Renmin University of China\\
         \textsuperscript{2}School of Information Technology and Management, University of International Business and Economics\\ 
         \{sunzhongxiang, kepuzhang\}@ruc.edu.cn, yu@uibe.edu.cn, \{wanghaoyu0924, junxu\}@ruc.edu.cn}

\abstract{
In this paper, we address the issue of using logic rules to explain the results from legal case retrieval. The task is critical to legal case retrieval because the users (e.g., lawyers or judges) are highly specialized and require the system to provide logical, faithful, and interpretable explanations before making legal decisions. Recently, research efforts have been made to learn explainable legal case retrieval models. However, these methods usually select rationales (key sentences) from the legal cases as explanations, failing to provide faithful and logically correct explanations. 
In this paper, we propose \textbf{N}eural-\textbf{S}ymbolic enhanced \textbf{L}egal \textbf{C}ase \textbf{R}etrieval (\textbf{NS-LCR}), a framework that explicitly conducts reasoning on the matching of legal cases through learning case-level and law-level logic rules. The learned rules are then integrated into the retrieval process in a neuro-symbolic manner. Benefiting from the logic and interpretable nature of the logic rules, NS-LCR is equipped with built-in faithful explainability. We also show that NS-LCR is a model-agnostic framework that can be plugged in for multiple legal retrieval models.
To showcase NS-LCR's superiority, we enhance existing benchmarks by adding manually annotated logic rules and introducing a novel explainability metric using Large Language Models (LLMs). Our comprehensive experiments reveal NS-LCR's effectiveness for ranking, alongside its proficiency in delivering reliable explanations for legal case retrieval.
 \\ \newline \Keywords{Legal Applications, Information Retrieval, Explainability} }

\begin{document}

\maketitleabstract

\section{Introduction}
\label{sec:intro}

Legal case retrieval retrieves relevant cases from a query and is a specialized Information Retrieval task. Due to its vital role in aiding legal practitioners, logical explanations for retrieved cases are essential. Only retrieved cases with accurate logical reasoning can serve as persuasive evidence for legal decisions~\cite{PRAKKEN2015214}.

\begin{figure}[h]
    \centering
    \includegraphics[width=\linewidth]{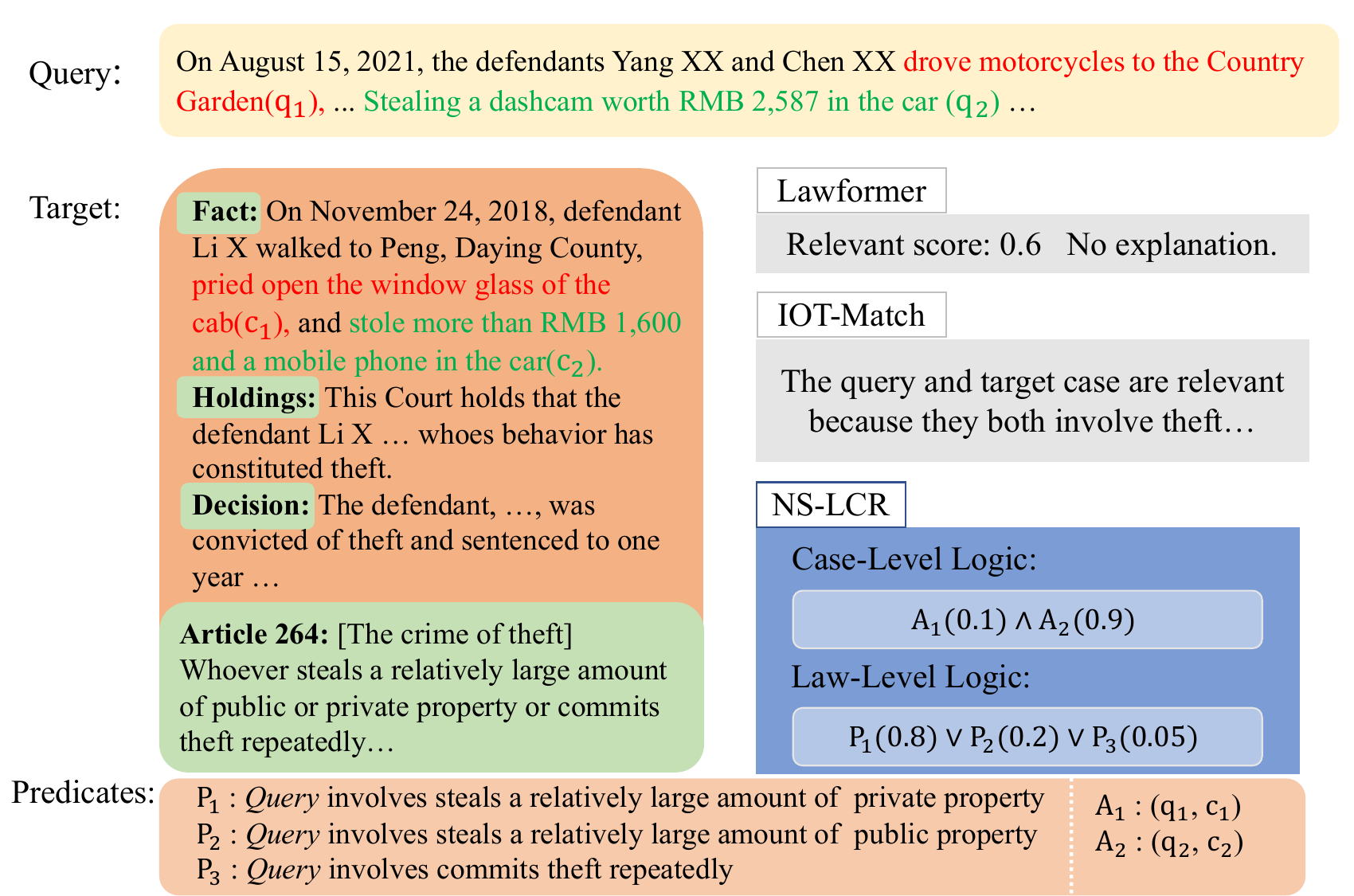}
    \caption{
  Explanations provided by different legal case retrieval models. Semantic models (e.g., Lawformer) only estimate the matching score.  Existing explainable methods (e.g., IOT-Match) provide sentences as explanations. NS-LCR aims to explain matching results with case and law logic rules. Article 264 of PRC Criminal Law, with three key facts \(P_{1}\), \(P_{2}\), and \(P_{3}\), applies to the target case.
    }
    \label{fig:intro}
\end{figure}

Deep learning advances have improved semantic retrieval of legal cases~\cite{shao2020bert,xiao2021lawformer,qin2023incorporating}. Most retrieval models focus on estimating the relevance scores of a target case given the query (shown as the Lawformer tab in Figure~\ref{fig:intro}). Additionally, in response to the need for explainability in legal case retrieval,~\citet{yu2022explainable} proposed IOT-Match, which generates explanations by extracting rationales (key sentences) from both query and target cases (shown as the IOT-Match tab in Figure~\ref{fig:intro}). However, IOT-Match cannot provide the users with an explicit logic reasoning process on whether the query and candidate cases are relevant or not. Furthermore, these explanations focus only on case facts, overlooking law articles' significance in assessing query and candidate case relevance~\cite{Sun2022LawAL}.

Recently, some studies have used logic for explanation. \citet{lee2022self} have demonstrated the effectiveness of learning rules from data for explanation. Furthermore, logic rule-based explanations surpass prior methods in human precision~\cite{alvarez2018towards}. \citet{ciravegna2023logic} proposed a unique type of concept-based neural network that provides first-order logic explanations for decision-making. 
Though these methods have shown promise in providing explanations for tasks in general domains, they cannot be directly adapted to legal retrieval because, in the legal domain, it is required that the judges make decisions based not only on case documents but also on law articles. Moreover, some studies highlight law articles' role in enhancing judgment prediction~\cite{zhong2018legal} and legal case matching~\cite{Sun2022LawAL}. How to incorporate the corresponding law articles in explicit logic reasoning is important while under-explored problems. It is expected that legal case retrieval models should explain decisions using logic rules from both cases and law articles, as shown in Figure~\ref{fig:intro}.



To tackle these issues, we proposes a model-agnostic framework called NS-LCR which learns logic rules from the query and target cases as the explanations for retrieved legal cases. Unlike studies that solely rely on text semantics for relevance scores, NS-LCR uses two neuro-symbolic modules to learn law-level and case-level logic rules.
Specifically, the law-level module forms first-order-logic (FOL) rules for each target case, extracting predicates from the case based on laws and connecting them with logic operations. Then, the legal relevance prediction is formalized as the fine-grained evaluation between the query and the FOL rule, which can be efficiently induced by fuzzy logic such as Łukasiewicz T-norm~\cite{klement2013triangular}. 
The case-level module forms the relevance rules by identifying the aligned sentences from the query and target cases. The learned relevance rules also work in a fuzzy logic fashion and provide evidence for the relevance prediction. 
Benefiting from these modules, NS-LCR not only provides the learned logic rules as explanations but also effectively improves the underlying retrieval model's performance in low-resource situations.
We take four well-known legal retrieval baselines as the underlying model of NS-LCR and conduct extensive experiments of high-resource and low-resource legal retrieval performance on LeCaRD~\cite{ma2021lecard} and ELAM~\cite{yu2022explainable}. To evaluate explanations, we use Large Language Models to check the effectiveness when applied to downstream tasks.

We summarize our contributions as follows:
(1) We analyze the importance of explicit logic reasoning in legal case retrieval. We further show that the law-level and case-level logic rules are critical in explaining the retrieved cases. 
(2) We propose a novel neural symbolic enhanced framework (\textbf{NS-LCR}) for explainable legal case retrieval by representing law articles and case documents into logic rules and involving the logic rules in legal case retrieval. 
(3)  We tested on two datasets and introduced a new \textbf{LLM-based} evaluation method. Results showed NS-LCR enhanced model performance and validated the importance of case and law-level rules in legal retrieval.


\section{Related Work}
\subsection{Legal Case Retrieval}
Traditional techniques emphasized legal issue decomposition and ontology construction~\cite{bench2012history, saravanan2009improving}, while recent advancements segment into network and text-based methods. Network-based strategies, such as the Precedent Citation Network (PCNet) and Hier-SPCNet, focus on citation clustering and domain encapsulation to assess case similarity~\cite{minocha2015finding, bhattacharya2020hier}. On the other hand, text-based methods leverage semantic analysis of case texts, with innovations like BERT-PLI, OPT-Match and Lawformer modeling paragraph interactions and specializing in case analysis~\cite{shao2020bert,yu-etal-2022-optimal,xiao2021lawformer}. Despite improvements, the explicability of these models remains a challenge, highlighted by the introduction of a tri-stage explainable model by~\citet{yu2022explainable, sun2023explainable}, which still falls short in logical reasoning~\cite{jain2019attention}.

\subsection{Logic as Explanation}

Studies indicate logic explains prediction results. Works are classified into concept-based explanations with predicates from concept set inputs, and data-based explanations that learn rules from data. \citet{barbiero2022entropy} used an entropy criterion to identify relevant concepts and extract First-Order Logic explanations from neural networks. \citet{ciravegna2023logic} introduced LENs that predict output concepts and provide First-Order Logic explanations based on input concepts. \citet{jain2022extending} enhanced LEN by testing perturbed input words on text classification. \citet{wu2021weakly} improved Natural Language Inference explainability by aligning detected phrases in sentences. Aligned units form weakly supervised logic reasoning. \citet{lee2022self} developed a framework using human priors to learn logic rules from data. Learned rules explain deep models’ output. \citet{feng2022neuro} presented a framework combining reinforcement learning and introspective revision for improved reasoning in natural language inference tasks.
Our study advances beyond existing approaches by focusing on the legal domain, leveraging the logical structure of the civil law system and the semi-structured nature of legal documents. We introduce the NS-LCR model, which generates more precise and detailed explanations, improving effectiveness in legal case retrieval compared to methods designed for general domains.

\section{Background and Preliminaries}
\subsection{Task Formulation}
\ywj{
Suppose that we have a set of collected samples $\mathcal{D}=\{(q, \mathcal{C}, \mathcal{L}, \mathcal{R})\}$. For each data instance, $q$ represents a query case submitted by the legal practitioner; $\mathcal{C}=\{c_1, c_2, \cdots, c_{N_C}\}$ represents a set of candidate cases (precedents) with size ${N_C}$ in which $c_i\in \mathcal{C}$ is potentially relevant to $q$ and thus support $q$'s legal judgement; $\mathcal{L}=\{l_1,l_2,\cdots,l_{N_L}\}$ represents the set of applicable laws with size ${N_L}$ that provides legal basis for the relevance judgement between $q$ and $c_i$; $\mathcal{R}$ represents the labeled ranking of $\mathcal{C}$ given the query case $q$.
NS-LCR aims at learning a ranking function $f: q\times \mathcal{C}\times\mathcal{L}\rightarrow \mathcal{R}\times \mathcal{E}$, where $\mathcal{E}$ denotes the desired logic explanations  corresponding to $\mathcal{R}$. 
}

As mentioned in~\autoref{sec:intro}, we consider two-level explanations for legal retrieval in this study, i.e., $\mathcal{E}=\{e_L,e_C\}$.
$e_L$ is a learned logic rule and denotes the law-level explanations that represent the alignment between $q$ and $l\in\mathcal{L}$ applicable for $c_i$.
$e_C$ is another learned logic rule and denotes the case-level explanations that represent the sentence-level alignment between $q$ and $c_i$.
To explicitly model the logic reasoning in legal retrieval, we represent $e_L$ and $e_C$ in the first-order-logic (FOL) format and evaluate them in the fuzzy logic way which we will introduce in the following sections.
Benefiting from the logic rules, NS-LCR not only provides explanations for the retrieval but also obtains law-level and case-level relevance scores respectively denoted as $r_L$ and $r_C$ by solving $e_L$ and $e_C$. 
NS-LCR further combines $r_L$, $r_C$, and semantic relevance score $r_N$ between $q$ and $c_i$ to rank candidates.

\subsection{Presenting law articles as FOL}
\label{sec:fol}


In this study, we present law articles in the FOL format. Specifically, we first manually extract $\mathcal{P}^i=\{P^i_1, P^i_2, \cdots, P^i_{N_P}\}$, a set of predicates  from each of $l_i\in\mathcal{L}$, where the predicate represents a key fact or a key circumstance~\cite{ma2021lecard}, ${N_P}$ denotes the number of predicates. Then, the extracted predicates are connected by logic operators, including conjunction ($\land$), disjunction ($\lor$), and negation ($\neg$) to form the clause. As a result, the FOL rules enables the precise expression of the relationships among all of the key facts and circumstances in a legal article, whereby such relationships denote the applicability of this law.
For example, as shown in Figure~\ref{fig:intro}, the predicates of ``Article 264: [The crime of theft] whoever steals a relatively large amount of public or private property or commits theft repeatedly'' include $P_1$ = ``steals a relatively large amount of private property'', $P_2$ = ``steals a relatively large amount of public property'', and $P_3$ = ``commits theft repeatedly''. Based on the relationship among predicates, we represent this law article as a FOL format logic rule $l:(P_1\lor P_2\lor P_3 \rightarrow Y)$, where $Y$ = ``crime of theft''.

\subsection{Fuzzy Logic}

\ywj{
In this study, as law articles are represented in the FOL format, NS-LCR learns the alignments between the query $q$ and the applicable law $l_i\in\mathcal{L}$ of the candidate case $c\in\mathcal{C}$ by the fuzzy logic. Specifically,
at the predicate level, NS-LCR determines whether $q$ satisfies $P\in\mathcal{P}^i$ of $l$ based on $q-P$ similarity, which we called \textbf{evaluation}. 
At the rule level, NS-LCR reasons whether $l_i$ is applicable for $q$
according to all of $q-P$ similarities in $l_i$, which we called \textbf{induction}. In other words, through fuzzy logic-based the evaluation and induction, NS-LCR computes the similarity between $q$ and $c$ with the guidance of $l_i$ since $c$ has already been judged in history and has a clear applicable law, which provides the law-level relevance measurement.}

\section{OUR APPROACH: NS-LCR}
\begin{figure*}
    \centering
\includegraphics[width=0.75\linewidth]{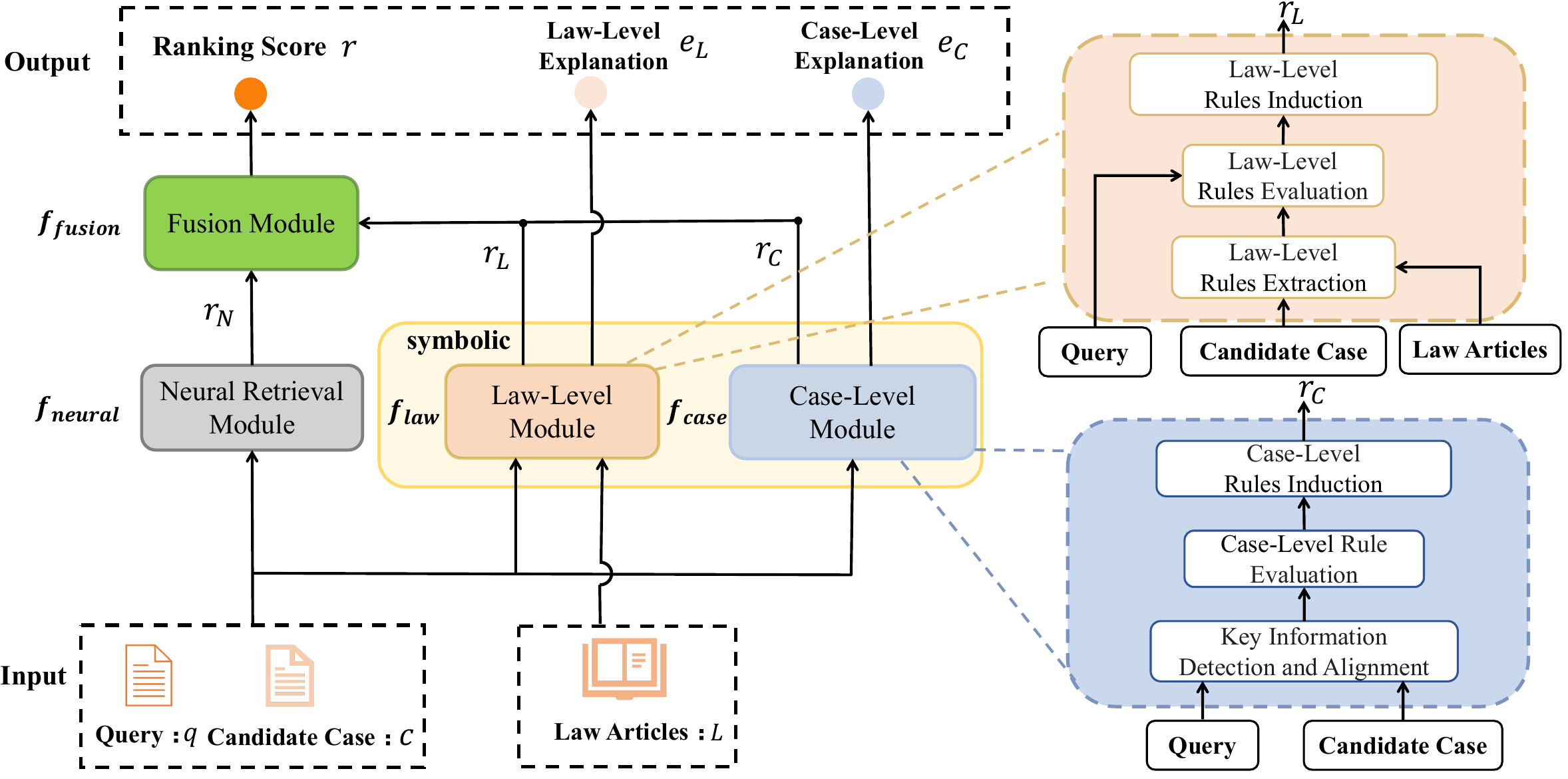}
    \caption{
    The overall architecture of the proposed model NS-LCR.}
    \label{fig:model_graph}
\end{figure*}
\subsection{General framework}

As illustrated in~\autoref{fig:model_graph}, our NS-LCR takes a query case $q$ and a candidate case $c$ as the input and predicts multi-level relevance scores along with two logic explanations $e_L$ and $e_C$. Specifically, NS-LCR achieves this goal through the following four modules:

\textbf{Neural retrieval module} $f_{neural}$ is responsible for predicting the relevance score from the semantic perspective given a pair of $(q,c)$:
\begin{equation}
\label{eq:neural}
   r_{N}=f_{neural}(q, c;\theta_{neural}),
\end{equation}
where $r_{N} \in \mathbb{R}$ denotes the degree of the relevance between $q$ and $c$; $\theta_{neural}$ denotes the learnable parameters in this module. Considering that NS-LCR is a general framework, $f_{neural}$ can be implemented by existing legal retrieval models, such as cross-encoder methods methods~\cite{xiao2021lawformer, shao2020bert} or dual-encoder methods~\cite{yu2022explainable, Sun2022LawAL}.

\textbf{Law-level module} $f_{law}$ incorporates law articles $\mathcal{L}$ in the form of FOL  into the input pair $(q,c)$ and outputs the law-level relevance score $r_L\in\mathbb{R}$ and the corresponding explanation $e_L$ in the FOL format:
\begin{equation}
\label{eq:law}
   (r_{L},e_{L})=f_{law}(q, c,\mathcal{L};\theta_{law}),
\end{equation}
where $\theta_{law}$ is the learnable parameters in this module, 
$f_{law}$ is implemented in a neuro-symbolic way,
which we will introduce the details in~\autoref{sec:law}.

\textbf{Case-level module $f_{case}$} is 
designed to learn the sentence-level alignment between $q$ and $c$. The module takes $(q,c)$ as the input and outputs the case-level relevance score $r_C\in\mathbb{R}$ and the corresponding explanation $e_C$ in the FOL format:
\begin{equation}
\label{eq:case}
   (r_{C},e_{C})=f_{case}(q, c;\theta_{case}),
\end{equation}
where $\theta_{case}$ is the learnable parameters in this module, $f_{case}$ is also implemented in a neuro-symbolic way, which we will introduce in~\autoref{sec:case}.

\textbf{Fusion module $f_{fusion}$} combines the outputs of all modules  $(r_N, r_L, r_C)$ to computes the final ranking score $r\in\mathbb{R}$ for the candidate case:
\begin{equation}
    r = f_{fusion}(r_N, r_L, r_C),
\end{equation}
where $f_{fusion}$ is the Weighted Reciprocal Rank Fusion (WRRF) to output $r$ by considering all three predicted ranks in a weighted manner:
\begin{equation}
\label{eq:fusion}
f_{fusion}=\sum_{i} \frac{w_{i}^{\pi}}{\epsilon+\pi(r_i)},
\end{equation}
where $\epsilon$ is the hyper-parameter for smoothness; $\pi(r_i)$ denotes the predicted rank of module $i\in\{f_{neural},f_{law},f_{case}\}$ based on the predicted relevance score $r_i$;  $w_i^{\pi}$ is a rank-aware module-specific weight that dynamically adjusts the importance among three types of relevance predictions:
\begin{eqnarray}
w_i^{\pi}=
\begin{cases}
1, & i=f_{neural},\\
\sin(\frac{\pi({r_i})}{\gamma}\times\frac{\pi}{2}) & i\in\{f_{law},f_{case}\},\\
\end{cases}
\end{eqnarray}
where $\gamma$ is a hyper-parameter and $\pi({r_i})$ is the predicted rank of the module $i$. 
Our Weighted Reciprocal Rank Fusion (WRRF) method enhances the traditional RRF by introducing dynamic weights in the ranking process, diverging from the uniform weighting strategy of RRF~\cite{cormack2009reciprocal, chen2022out}. WRRF prioritizes the neural retrieval module for higher-ranked predictions, where it is most confident, and increases the weight of the symbolic module (targeting law and case-specific information) for lower-ranked items to improve accuracy where neural confidence wanes.

\subsection{Law-level Module}
\label{sec:law}
In this section, we introduce the law-level module $f_{\text{law}}$ to model the relevance between a query case and candidates, guided by law articles, as depicted in the top-right of~\autoref{fig:model_graph}.
\subsubsection{Law-level Rule Evaluation}

Formally, suppose there is a query-candidate pair $(q,c)$ and a set of FOL format law articles $L\subset\mathcal{L}$ corresponding to $c$~\footnote{Each candidate case in civil law system is associated with relevant judged law articles, allowing Law-Level
Rules Extraction to be efficiently executed using straightforward text processing techniques.}.  For each of $l_i\in L$ which is represented as a connective of $N_P$ predicates $\mathcal{P}^i=\{P^i_1, P^i_2, \cdots, P^i_{N_P}\}$ as mentioned in~\autoref{sec:fol},
we first measure the alignment between $q$ and $P^i_j\in\mathcal{P}^i$ by:
\begin{equation}
\label{eq:predicate}
    s_{P^i_j}=f_P(q,P^i_j),
\end{equation}
where $P^i_j$ denotes the predicate which represents a fact or a circumstance of $l_i$; $s_{P^i_j}\in [0, 1]$ is the $(q,P_i^j)$ relevance score representing the degree to which $q$ satisfies $P^i_j$; $f_P$ is implemented by a pre-trained language model (PLM) trained on a large Chinese criminal judgment corpus\footnote{The corpus is from LeCaRD~\cite{ma2021lecard}. We excluded data from the test set to prevent
data leakage}. Specifically, we construct the input of the PLM in the form of [CLS] + $P^i_j$ + [SEP] + $q$ + [SEP]". The derived embedding of "[CLS]" token in the last layer is then fed to an MLP to compute the score $s_{P^i_j}$. 

\subsubsection{Law-level Explanation and Induction}
Given all of the alignment scores between $q$ and each of $L\subset\mathcal{L}$ corresponding to $c$, we represent the law-level explanation $e_L$ for a pair $(q, c)$ as: 
\begin{equation}
    e_L=\bigvee_{i=1}^{N_L}\bigwedge_{j=1}^{N_P}P^i_j\oplus s_{P^i_j},
\end{equation}
where $\oplus$ denotes concatenation operation; $N_L$ and $N_P$ respectively denote the number of applicable laws to $c$ and the number of predicates in law $l_i$; $\wedge$ and $\vee$ respectively denotes the conjunction and disjunction operations.
The derived explanation $e_L$ precisely indicates the similarity between query $q$ and all predicates (facts or circumstance) in the law article $l_i$ applicable to candidate case $c$.
Taking~\autoref{fig:intro} as an example, the
law-level explanation $e_L=P_1(0.8)\vee P_2(0.2)\vee P_3(0.05)$ means the query respectively has the relevance score $0.8$, $0.2$, $0.05$ to the predicate $P_1$ = ``steals a relatively large amount of private property'', predicate $P_2$ = ``steals a relatively large amount of public property'', and predicate $P_3$ = ``commits theft repeatedly''.

To further induce the law-level relevance score of $(q,c)$, we combine all query-predicates alignment scores across laws using T-norm fuzzy logics~\cite{gottwald2005triangular}:
\begin{equation}
r_{L} =
\frac{\sum_{i=1}^{N_L}\Gamma (\wedge_{j=1}^{N_P}s_{P^{i}_j})}{N_L},
\end{equation}
where $N_L$ and $N_P$ respectively denote the number of applicable laws to $c$ and the number of predicates in law $l_i$, $\Gamma()$ denotes the
Łukasiewicz t-norm~\cite{klement2013triangular,li2019augmenting} that maps the discrete outputs into continuous real values to achieve the induction.
 As our law-level logic rules are expressed in conjunctive normal form, we can compute $r_L$ by the following steps: (1) use $\Gamma\left(\bigwedge_i P_i\right)$ (or $\Gamma\left(\neg \bigvee_i P_i\right))$ to aggregate predicate-level predictions and get the score for each law $l_{i}$;
(2) combine all law-level predictions from (1) to get the final score $r_L$.
Please note that in real legal practice, a law may include multiple circumstances connected by $\vee$, indicating that any of the circumstances being satisfied is sufficient to establish the charge. Therefore, step (1) may be computed several times and connected by using $\Gamma\left(\bigvee_i P_i\right)$ (or $\Gamma\left(\neg \bigwedge_i P_i\right))$.

\subsection{Case-level Module}
\label{sec:case}
The law-level module provides a relevance measure rooted in law articles, yet due to legal cases' semi-structured feature, our legal retrieval model incorporates a case-level module \(f_{\text{case}}\) for fine-grained sentence alignment between queries and candidates, leveraging logic rules modeling as shown in~\autoref{fig:model_graph}.

\ywj{
\subsubsection{Case-level Rule Evaluation} 
Formally, given a query-candidate pair $(q,c)$, 
we split $q$ and $c$ into individual sentences respectively denoted by $\left\{x_i\right\}_{i=1}^{N_q}$ and $\left\{y_j\right\}_{j=1}^{N_c}$ and use use a pre-trained Sentence-BERT~\cite{reimers2019sentence} to extract the corresponding embeddings respectively denoted as $\{\mathbf{x}_i\}_{i=1}^{N_q}$, $\{\mathbf{y}_j\}_{j=1}^{N_c}$, where $N_q$ and $N_c$ is the number of sentences in $q$ and $c$.
For each sentence $\mathbf{x}_{i}$, we seek the $K$ most similar sentences from $\{\mathbf{y}\}_{j=1}^{N_c}$ based on their embeddings, where $K$ is a small value. The case-level logic rule can be constructed as follows:
\begin{equation}
e_C=\bigwedge_{i=1}^{N_q}\bigwedge_{j=1}^{K}(x_i,y_j)\oplus \cos(\mathbf{x}_i,\mathbf{y}_j),
\end{equation}
where $\cos(\cdot)$ denotes the cosine similarity function~\footnote{The sentence pairs of cosine similarity $\leq 0$ are filtered.}. The conjunctive form ensures that learning $e_C$ is a way to help identify relevant facts and circumstances from $(q,c)$. Meanwhile considering $e_C$ is constructed with a small portion of sentences from $(q,c)$, it naturally filters the noise from  $(q,c)$.
}
\subsubsection{Case-level Explanation and Induction}
Given extracted sentence-pair similarity predictions, we induce the case-level relevance score $r_C$ by:
\begin{equation}
r_C=\left\{\prod_{i=1}^{N_q}\prod_{j=1}^K\cos(\mathbf{x}_i,\mathbf{y}_j)\right\}^{\frac{1}{N_q*K}}.
\end{equation}
For simplicity, we directly apply the geometric mean to aggregate all sentence pair predictions due to its inclination towards low scores --- if there is at least one pair with a low similarity score, then $r_{C}$ is also low. It expects $r_C$ to illustrate the similarity between all the facts and circumstances in $(q,c)$.

\subsection{Model Training}

We pre-train $\theta_{law}$ using the following steps:
(1) For each candidate, we create a pseudo query $\Tilde{q}$ from its basic fact description;
(2) Using BM25~\cite{robertson1995okapi}, we select relevant predicates $\Tilde{p}^+$ from law articles, marking them as positive (label 1);
(3) We also sample negative examples, distinguishing between hard and easy negatives based on chapters~\footnote{
In the civil law system, law articles within the same chapters are similar, while those from different chapters are dissimilar.}. Predicates $\Tilde{p}^-$ from these are sampled with balanced ratios and labeled as 0;
(4) Using the BCE loss function, we pre-train $f_{law}$:
$$
\ell_{law}=-\frac{1}{n}\sum_{i}^{N}(y_i\log(p_i)+(1-y_i)\log(1-p_i)),
$$
where $y_i$ is the predicate's label, and $p_i=f_{law}(\Tilde{q}_i,\Tilde{p}_i,\mathcal{L})$.
After this, $\theta_{law}$ remains fixed during other training and inference stages.

For $\theta_{case}$, we employ Sentence-BERT~\cite{reimers2019sentence} directly without adjustments.

For $\theta_{neural}$, we train the Neural Retrieval Module using the MSE loss function comparing predicted $\hat{r}_N$ and actual relevance scores $r$ for query-candidate pairs:
$ \ell= \frac{1}{n}\sum_{i=1}^{N}(\hat{r}_N^{i} - r^{i})^2 .$

\section{Experiments}

\begin{table*}[!ht]
\caption {Performance comparisons between NS-LCR and the baselines. The boldface represents the best performance. `$\dagger$' indicates that the improvements over all of the baselines are statistically significant (t-tests, $p\textrm{-value}< 0.05$). 
}\label{tab:main results} 
\centering
\resizebox{0.95\linewidth}{!}{
\begin{tabular}{lcccccccccccc}
\toprule 

\multirow{2}{*}{\textbf{Models}} & \multicolumn{6}{c}{\textbf{LeCaRD}} & \multicolumn{6}{c}{\textbf{ELAM}}                                                                      \\
\cmidrule(lr){2-7}\cmidrule(lr){8-13}
  & \textbf{P@5}& \textbf{P@10}& \textbf{MAP} &  \textbf{NDCG@10}    &  \textbf{NDCG@20} & \textbf{NDCG@30} & \textbf{P@5}& \textbf{P@10}& \textbf{MAP} &  \textbf{NDCG@10}    &  \textbf{NDCG@20} & \textbf{NDCG@30} \\

\cdashline{1-13}
Criminal-BERT &0.430    &0.425 &0.498  &0.728  &0.760 &0.799  &0.541 &0.512  &0.598  &0.641  &0.778  &0.821   \\
Cat-Law(Criminal-BERT) &0.410    &0.405 &0.497  &0.745 &0.767 &0.817  &0.447 &0.535  &0.547  &0.603  &0.745  &0.780  \\
EPM(Criminal-BERT) &0.410 &0.415 &0.518  &0.740  &0.782 &0.813  &0.565 &0.541  &0.613  &0.633  &0.778  &0.804  \\
Law-Match(Criminal-BERT)& 0.420&0.433 & 0.502& 0.746& 0.778& 0.815& 0.626& 0.542& 0.628& 0.657& 0.791& 0.829\\
NS-LCR(Criminal-BERT) &\textbf{0.440}$^\dagger$   &\textbf{0.440}$^\dagger$ &\textbf{0.566}$^\dagger$  &\textbf{0.758}$^\dagger$ &\textbf{0.793}$^\dagger$ &\textbf{0.833}$^\dagger$ &\textbf{0.647}$^\dagger$ &\textbf{0.553}$^\dagger$ &\textbf{0.644}$^\dagger$ &\textbf{0.688}$^\dagger$  &\textbf{0.830}$^\dagger$  &\textbf{0.857}$^\dagger$ \\
\cdashline{1-13}
Lawformer &0.460    &0.425 &0.497  &0.739  &0.759 &0.800  &0.588 &0.582  &0.612  &0.647  &0.775  &0.801   \\
Cat-Law(Lawformer) &0.460    &0.405 &0.464  &0.690  &0.722 &0.764  &0.576 &0.559  &0.618  &0.643  &0.749  &0.810   \\
EPM(Lawformer) &0.450    &0.415 &0.469  &0.695  &0.732 &0.782  &0.553 &0.576  &0.611  &0.640  &0.771  &0.810  \\
Law-Match(Lawformer) & 0.450& 0.432& 0.496& 0.768& 0.773& 0.819& 0.634& 0.578& 0.644& 0.654& 0.790& 0.818\\
NS-LCR(Lawformer) &\textbf{0.480}$^\dagger$  &\textbf{0.450}$^\dagger$ &\textbf{0.531}$^\dagger$ &\textbf{0.786}$^\dagger$ &\textbf{0.810}$^\dagger$ &\textbf{0.841}$^\dagger$ &\textbf{0.635} &\textbf{0.594}$^\dagger$ &\textbf{0.656}$^\dagger$ &\textbf{0.701}$^\dagger$ &\textbf{0.829}$^\dagger$ &\textbf{0.859}$^\dagger$ \\    
 \cdashline{1-13}
Bert-PLI &0.420  &0.455 &0.490 &0.781  &0.821 &0.868 &0.506 &0.524 &0.580  &0.598  &0.772  &0.794   \\
Cat-Law(BERT-PLI) &0.420    &0.420 &0.490  &0.773  &0.817 &0.861  &0.518 &0.535&0.588  &0.636  &0.777  &0.811   \\
EPM(BERT-PLI) &0.430   &0.425 &0.486  &\textbf{0.793}&0.823 &0.874  &0.541 &0.524  &0.612  &0.635  &0.796  &0.827  \\
Law-Match(BERT-PLI) &\textbf{0.450} & 0.425& 0.519& 0.789& 0.829& 0.868& 0.600&0.556 & 0.629 &0.687 & 0.823& 0.843\\
NS-LCR(BERT-PLI) &\textbf{0.450}   &\textbf{0.460}$^\dagger$ &\textbf{0.544}$^\dagger$ &0.792 &\textbf{0.831} &\textbf{0.876} &\textbf{0.635}$^\dagger$ &\textbf{0.576}$^\dagger$ &\textbf{0.643}$^\dagger$&\textbf{0.688} &\textbf{0.836}$^\dagger$ &\textbf{0.858}$^\dagger$ \\
\cdashline{1-13}
BERT-ts-L1 &0.430    &0.425 &0.516  &0.736  &0.776 &0.819  &0.518 &0.529  &0.577 &0.613  &0.759  &0.795   \\
Cat-Law(BERT-ts-L1) &0.430    &0.405 &0.508  &0.733  &0.754 &0.805  &0.459 &0.535  &0.544  &0.601  &0.743  &0.778   \\
EPM(BERT-ts-L1) &0.420    &0.420 &0.520  &0.744  &0.790 &0.815  &0.553 &\textbf{0.582}  &0.595  &0.631  &0.759  &0.797  \\
Law-Match(BERT-ts-L1) &0.430 & 0.427& 0.528& 0.759&0.793 & 0.829& 0.529& 0.494& 0.591& 0.609& 0.749& 0.807\\
NS-LCR(BERT-ts-L1) &\textbf{0.440}$^\dagger$  &\textbf{0.440}$^\dagger$ &\textbf{0.550}$^\dagger$&\textbf{0.771}$^\dagger$&\textbf{0.798}$^\dagger$ &\textbf{0.847}$^\dagger$ &\textbf{0.565}$^\dagger$ &0.576 &\textbf{0.640}$^\dagger$ &\textbf{0.687}$^\dagger$ &\textbf{0.821}$^\dagger$ &\textbf{0.852}$^\dagger$ \\
\bottomrule
\end{tabular}
}
\end{table*}

\subsection{Experimental settings}

\subsubsection{\textbf{Dataset}} 
\label{sec:data}

\begin{table}[t]
\small
 \centering
 \caption{Statistics of the annotated FOL rules of law articles.}
\begin{tabular}{ccccc}
\toprule
 \multirow{2}{*}{total articles} & \multicolumn{4}{c}{ FOL } \\
\cline {2-5}
 & $\neg$ & $\wedge$ & $\vee$ & $\rightarrow$ \\
  \midrule
 441&44 & 2259& 1625 &1232 \\

\bottomrule
\end{tabular}
\label{tab:fol_article_analysis}
\end{table}

\begin{table}[t]
\small
    \centering
      \caption{Statistics of LeCaRD and ELAM}
      \resizebox{0.99\linewidth}{!}{
    \begin{tabular}{l|c|c}
    \toprule
        ~ & LeCaRD & ELAM  \\ \midrule
        \# total queries & 107 & 85  \\ 
        \# candidate cases per query & 100 & 50  \\ 
        avg. \# relevant cases per query & 10.33 & 9.06  \\ 
        avg. \# sentences per query & 6.91 & 16.24  \\ 
        avg. \# sentences per candidate case  & 86.63 & 43.43  \\ 
        avg. \# cited law articles per candidate case & 6.5 & 5.62 \\ 
        \bottomrule
    \end{tabular}}
  
    \label{tab:data_analysis}
\end{table}

The experiments\footnote{Both the source code and dataset are available at:~\url{https://github.com/ke-01/NS-LCR}.} were conducted based on two
publicly available datasets: ELAM~\cite{yu2022explainable} and LeCaRD~\cite{ma2021lecard}.

\textbf{LeCaRD} is a legal case retrieval dataset that contains 107 queries and 43,000 target cases~\footnote{Some experiment results are differences from Lecard's paper due to code errors in the original LeCaRD implementation, rectified in commit 89b7bf8.}. For each query, 100 target cases are provided, each assigned a 4-level relevance label. All criminal cases were published by the Supreme People's Court of China. 

\textbf{ELAM} is an explainable legal case matching dataset, which contains 5000 pairs of annotated cases, and each pair is manually assigned a matching label which is either match (2), partially match (1), or mismatch (0). We transformed this dataset into a legal case retrieval dataset by following the steps: In formulating queries, we capitalize on the paired cases and labels present in the ELAM dataset, enumerating the frequency of each case among these pairs. Instances with a prevalence surpassing six occurrences are designated as queries within match-labeled pairs, culminating in 85 queries. For each specific query, the initially constructed candidate set comprises the cases paired with it in ELAM. Simultaneously, supplementary cases are integrated to expand the candidate pool's dimensions. Considering ELAM's limited data volume, a subset of cases is chosen from LeCaRD's candidate set to establish a corpus. In adherence to LeCaRD's methodology for assembling a candidate pool, we employ a triad of retrieval models BM25~\cite{robertson1995okapi} and TF-IDF~\cite{salton1988term} to obtain the top 100 cases from the corpus for each query. Cases that secure a position within the top 100 in a minimum of one model are assimilated as hard negatives into the corresponding candidate set. In contrast, those absent from the top 100 across all three models are annexed as soft negatives. With an approximate 1:4 ratio between hard and soft negatives, the outcome for each query consists of a corresponding set of 50 candidate cases.

We normalize the relevance labels in LeCaRD and ELAM, which have multiple levels, to ensure that the ground-truth relevance scores range from 0 to 1. Table~\ref{tab:data_analysis} lists the statistics of the datasets.

Since both datasets correspond to Chinese legal case retrieval and NS-LCR targets learning law-level logic rules, we request three annotators manually label the articles of Criminal Law of the People's Republic of China
 in the FOL format, as~\autoref{sec:fol} mentioned.
The annotators are postgraduate students in artificial intelligence who have undergone legal training, and the annotation results undergo double-checking by professionals. Some basic statistics of the $L$ are listed in~\autoref{tab:fol_article_analysis}.

\subsubsection{\textbf{Baselines}}
To verify the effectiveness of NS-LCR, we apply it to the following underlying models:

\textbf{Criminal-BERT}~\cite{zhong2019openclap} is a legal domain pre-trained model based on BERT~\cite{devlin2018bert}, which is fine-tuned on millions of criminal legal document. The cross-encoder architecture has been utilized to predict the relevance between the query and the candidate case.

\textbf{Lawformer}~\cite{xiao2021lawformer} is a Longformer-based pre-trained language model training millions of Chinese legal cases to represent long legal documents better. In the experiment, we concatenate the input cases to Lawformer and use the mean pooling of Lawformer’s output to conduct matching.

\textbf{BERT-PLI}~\cite{shao2020bert} uses BERT to capture the semantic relationships at the paragraph level. Then it uses RNN and Attention model to infer the relevance between the two cases. Finally, it uses an MLP to calculate the aggregated embeddings similarity score.

\textbf{BERT-ts-L1}~\cite{shao2022understanding} optimizes the Criminal-BERT's attention weights with the attention of users majoring in law for relevance prediction in legal case retrieval. The optimized model is then used to predict the relevance of the query and candidate case.

We also compare NS-LCR with three baselines that can also both consider the case-level and law-level in models. The first is an intuitive baseline that appends the contents of cited law articles to the original cases, forming new extended legal cases. Existing matching models of Criminal-BERT, Lawformer, BERT-PLI, and BERT-ts-L1 can be applied to the extended legal cases, denoted as \textbf{Cat-Law (Criminal-BERT)}, \textbf{Cat-Law (Lawformer)}, \textbf{Cat-Law (BERT-PLI)}, and \textbf{Cat-Law (BERT-ts-L1)}, respectively. 
The second baseline is EPM~\cite{feng-etal-2022-legal}, which employs an attention mechanism to incorporate article semantics into the legal judgment prediction models. Existing matching models of Sentence-BERT, Lawformer, BERT-PLI, and BERT-ts-L1 can be applied to EPM, denoted as \textbf{EPM (Criminal-BERT)}, \textbf{EPM (Lawformer)}, \textbf{EPM (BERT-PLI)}, and \textbf{EPM (BERT-ts-L1)}, respectively. The third baseline is Law-Match~\cite{Sun2022LawAL}, which learns legal case retrieval models by respecting the corresponding law articles as instrumental variables~(IVs) and legal cases as treatments. Then, IV decomposition and recombination are used to learn the legal case embedding. Law-Match is model-agnostic and can apply to Sentence-BERT, Lawformer, BERT-PLI, and BERT-ts-L1, denoted as \textbf{Law-Match~(Criminal-BERT)}, \textbf{Law-Match~(Lawformer)}, \textbf{Law-Match~(BERT-PLI)}, and \textbf{Law-Match~(BERT-ts-L1)}, respectively.

The proposed NS-LCR is also model-agnostic. In the experiments, we applied NS-LCR to the underlying models of  Criminal-BERT, Lawformer,  BERT-PLI, and BERT-ts-L1, achieving four versions and referred to as \textbf{NS-LCR (Criminal-BERT)}, \textbf{NS-LCR (Lawformer)},  \textbf{NS-LCR (BERT-PLI)}, and \textbf{NS-LCR (BERT-ts-L1)} respectively.

\subsection{\textbf{Implementation Details}}
\label{app:Implementation Details}
We optimize the hyperparameters of NS-LCR's base models through grid search on the validation set, employing Adam~\cite{kingma2015method}. The batch size is selected from \{2, 8, 16\}, while the learning rate is chosen from \{2e-5, 3e-6\}. The other parameters of the base models remain in line with their original paper.
In the law-level module, we fine-tune Criminal-BERT with a batch size of 24 and a learning rate 2e-5 to acquire the Predicate Evaluation Module. 
As for the case-level module, we tune $K$ from $\{1, 3, 5\}$. 
In terms of the fusion module, we set the hyper-parameter $\epsilon=60$ for smoothness and tune $\gamma\in\{0, 1, 2, 50\}$~\footnote{$\gamma=0$ represents equal weights assigned to ${f_{neural},f_{law},f_{case}}$ outputs.} to balance the effects of the three modules.

\subsection{{Evaluate Explainable Legal Case Retrieval}}

In assessing retrieval accuracy, we adopt precision metrics like P@5, P@10, MAP, and ranking metrics such as NDCG@10, NDCG@20, and NDCG@30, as per~\cite{ma2021lecard}.

We introduce a novel explanation evaluation method with help of large language models (LLMs).
Leveraging retrieval argument LLMs allows updated knowledge acquisition, improving generation~\cite{mialon2023augmented}. Believing explainable retrieval assists in comprehension, we posit that explanations can aid LLMs in downstream tasks. Thus, we transformed ELAM and LeCaRD dataset queries into a legal judgment task for LLMs, using four prompt types, including zero-shot and few-shot prompts with/without explanations~\cite{brown2020language,sun2023short}~\footnote{Explanations, based on logic rules, exclude predicates with scores below 0.5. Logic operators are translated to natural language: 'and' for $\land$, 'or' for $\lor$, 'not' for $\neg$. The prompts were translated to Chinese.
}:
    \begin{shaded}
    \noindent \footnotesize{\textbf{Zero-shot ~\textcolor[RGB]{255,0,0}{with}/without explanation: }}
    
    \noindent \footnotesize{Please answer the criminal name for the query fact description based on the relevant cases.} 
    
    \noindent \footnotesize{The query is~\textcolor[RGB]{101,42,150}{{[fact description]}}. }
    
    \noindent \footnotesize{Evidence:~\textcolor[RGB]{101,42,150}{{[relevant case]}}+~\textcolor[RGB]{255,0,0}{[explanation]}.  }
    
    \noindent \footnotesize{The answer is:}
    \end{shaded}

    \begin{shaded}
    \noindent \footnotesize\textbf{Few-shot ~\textcolor[RGB]{255,0,0}{with}/without explanation: }
    
    \noindent \footnotesize{Please answer the criminal name for the query fact description based on the relevant cases.} 

    \noindent \footnotesize{The query is~\textcolor[RGB]{101,42,150}{{[fact description]}}. }

    \noindent \footnotesize{Evidence:~\textcolor[RGB]{101,42,150}{{[relevant case]}}+~\textcolor[RGB]{255,0,0}{[explanation]}.  }

    \noindent \footnotesize{The answer is: ~\textcolor[RGB]{101,42,150}{{[criminal name]}}. }
    
    \noindent \footnotesize{The query is~\textcolor[RGB]{101,42,150}{{[fact description]}}. }
    
    \noindent \footnotesize{Evidence:~\textcolor[RGB]{101,42,150}{{[relevant case]}}+~\textcolor[RGB]{255,0,0}{[explanation]}. } 
    
    \noindent \footnotesize{The answer is:}
    \end{shaded}

Explanation quality is determined by LLM prediction accuracy for various prompts.

We employ \textbf{text-davinci-003}~\cite{ouyang2022training} and \textbf{gpt-3.5-turbo}. With the temperature set to $0$, we chose one relevant case per prompt due to model length constraints.

\subsection{Experimental results and analysis}


\subsubsection{\textbf{Comparison against baselines.}}

Based on results in~\autoref{tab:main results}, NS-LCR variants (including Criminal-Bert, Lawformer, Bert-PLI, BERT-ts-L1) surpassed their neural retrieval counterparts in six metrics on LeCaRD and ELAM with significance (t-tests, $p$-value <0.05). This underscores the symbolic module's potency in melding law article knowledge via law-level logic rules and enhancing retrieval by extracting key details from cases using case-level logic rules. Additionally, NS-LCR outperforms frameworks like Cat-Law, EPM, and Law-Match, highlighting the superiority of integrating detailed legal behaviors through logic rules over direct embedding encoding.



\subsubsection{\textbf{Quality of Logic Rules as Explanations.}}
        
        

\begin{table}[t]
    \centering
      \caption{Performance of LLMs across different  prompts on legal judgement prediction on LeCaRD.}
      \resizebox{0.95\linewidth}{!}{
    \begin{tabular}{llc}
    \toprule
        LLMs & Prompt & Accuracy \\ \midrule
         \multirow{4}{*}{\textbf{text-davinci-003 }} & Zero-shot w/o explanation & 0.621 \\ 
        ~ & Zero-shot w/ explanation & \textbf{0.656} \\ 
        \cdashline{2-3}
         ~ & Few-shot w/o explanation & 0.652  \\ 
        ~ & Few-shot w/ explanation & \textbf{0.707} \\ 
        \midrule
        
        \multirow{4}{*}{\textbf{gpt-3.5-turbo}} & Zero-shot w/o explanation & 0.675\\ 
        ~ & Zero-shot w/ explanation & \textbf{0.769} \\ 
        \cdashline{2-3}
        ~ & Few-shot w/o explanation & 0.707 \\ 
        ~ & Few-shot w/ explanation &  \textbf{0.832} \\   
        \bottomrule
    \end{tabular}}
    \label{tab:data_exp_lecard}
\end{table}

\begin{table}[t]
    \centering
      \caption{Performance of LLMs across different  prompts on legal judgement prediction on ELAM.}
    \resizebox{0.95\linewidth}{!}{
    \begin{tabular}{llc}
    \toprule
        LLMs & Prompt & Accuracy \\ \midrule
         \multirow{6}{*}{\textbf{text-davinci-003 }} & Zero-shot w/o explanation &0.672  \\ 
        ~ & Zero-shot w/ explanation~(IOT-Match) &0.841  \\
        ~ & Zero-shot w/ explanation~(NS-LCR) &\textbf{0.927}  \\
        \cdashline{2-3}
         ~ & Few-shot w/o explanation &0.782   \\ 
         ~ & Few-shot w/ explanation (IOT-Match) &0.853  \\
        ~ & Few-shot w/ explanation (NS-LCR)&\textbf{0.951}  \\ 
        \midrule
        
        \multirow{6}{*}{\textbf{gpt-3.5-turbo}} & Zero-shot w/o explanation &0.780  \\ 
        ~ & Zero-shot w/ explanation~(IOT-Match) &0.876  \\
        ~ & Zero-shot w/ explanation~(NS-LCR) &\textbf{0.916}  \\
        \cdashline{2-3}
         ~ & Few-shot w/o explanation &0.794   \\ 
         ~ & Few-shot w/ explanation (IOT-Match) &0.888  \\
        ~ & Few-shot w/ explanation (NS-LCR)&\textbf{0.969}  \\ 
        \bottomrule
    \end{tabular}
    }
    \label{tab:data_exp_elam}
\end{table}



Tables \ref{tab:data_exp_lecard} and \ref{tab:data_exp_elam} show outcomes from logic explanations on LeCaRD and ELAM. We infer: 
(1) Across both LLMs (text-davinci-003 and gpt-3.5-turbo) and datasets (LeCaRD and ELAM), evidence with cases and explanations outperformed evidence with only cases for both zero-shot and few-shot prompts. This highlights the significance of explainable legal case retrieval models in boosting human understanding and LLMs' comprehension of relevant cases; 
(2) \autoref{tab:data_exp_elam} contrasts IOT-Match~\cite{yu2022explainable} and NS-LCR explanations on ELAM\footnote{We report only on ELAM due to IOT-Match's need for explanation labels present only in ELAM.}. NS-LCR explanations were found superior, indicating their high-quality nature.



\subsubsection{\textbf{Ablation study.}}

To evaluate the effectiveness of NS-LCR's symbolic components, we conducted an ablation study on LeCaRD\footnote{We only focused on LeCaRD due to ELAM's adaptation to the legal case retrieval dataset.\label{fn:only_lecard}}. Our integrations included: (1) $+$ \textbf{Law-level Module}, (2) $+$ \textbf{Case-level Module}, and (3) $+$ \textbf{NS-LCR} (combining both modules) into the base model. As shown in \autoref{tab:ablation}, both modules significantly boosted retrieval metrics, with the Law-level Module incorporating FOL-form law articles, and the Case-level Module applying logic rules from queries and cases. The combined NS-LCR framework markedly surpassed the base model, highlighting the importance of merging law and case information via logic rules for neural retrieval model efficacy. Notably, the Law-level module's impact waned at higher $K$ values, especially alongside the Case-level module, suggesting its optimality for complex differentiation tasks may diminish with broader top-ranking item arrays, potentially overshadowing its utility.

\begin{table}[t]
\caption { Ablation study of NS-LCR on LeCaRD.
}\label{tab:ablation} 
\centering
\resizebox{\linewidth}{!}{
\begin{tabular}{lcccccc}
\toprule

  \textbf{Models} & \textbf{P@5}& \textbf{P@10}& \textbf{MAP} &  \textbf{NDCG@10}    &  \textbf{NDCG@20} & \textbf{NDCG@30} \\
\midrule
\cdashline{1-7}
Criminal-BERT &0.430  &0.425  &0.498  &0.728 &0.760  &0.799   \\
$+$ Law-level Module &0.430  &0.430  &0.530  &0.734 &0.767  &0.805   \\
$+$ Case-level Module &0.430  &0.425  &0.506  &0.741 &\textbf{0.793} &\textbf{0.844}  \\
$+$ NS-LCR &\textbf{0.440}  &\textbf{0.440}  &\textbf{0.566}  &\textbf{0.758} &\textbf{0.793} &0.833  \\
\cdashline{1-7}
 Lawformer &0.460  &0.425  &0.497  &0.739 &0.759  &0.800   \\
$+$ Law-level Module &\textbf{0.480}  &0.435  &0.517  &0.772 &0.773  &0.812   \\
$+$ Case-level Module&0.470  &0.445  &0.508  &0.766 &0.802  &\textbf{0.849}  \\
$+$ NS-LCR &\textbf{0.480 } &\textbf{0.450}  &\textbf{0.531}  &\textbf{0.786} &\textbf{0.810} &0.841  \\   
 \cdashline{1-7}
BERT-PLI &0.420  &0.455  &0.490  &0.781 &0.821  &0.868   \\
$+$ Law-level Module&0.440  &0.455  &0.510  &0.782 &0.823  &0.868   \\
$+$ Case-level Module&0.430  &0.455  &0.492  &0.784 &0.827  &0.872  \\
$+$ NS-LCR &\textbf{0.450}  &\textbf{0.460}  &\textbf{0.544} &\textbf{0.792} &\textbf{0.831 }&\textbf{0.876}  \\  
\cdashline{1-7}
BERT-ts-L1 &0.430  &0.425  &0.516  &0.736 &0.776  &0.819   \\
$+$ Law-level Module&0.430  &0.435  &0.545  &0.755 &0.780  &0.823   \\
$+$ Case-level Module&0.430  &0.430  &0.526  &0.743 &0.787  &\textbf{0.858}  \\
$+$ NS-LCR &\textbf{0.440}  &\textbf{0.440}  &\textbf{0.550} &\textbf{0.771} &\textbf{0.798}  &0.847  \\  
\bottomrule
\end{tabular}}
\end{table}
\subsubsection{\textbf{Low-resource scenarios.}}

\begin{figure}[t]
  \centering
  
  \begin{minipage}[t]{0.49\linewidth}
    \includegraphics[width=\linewidth]{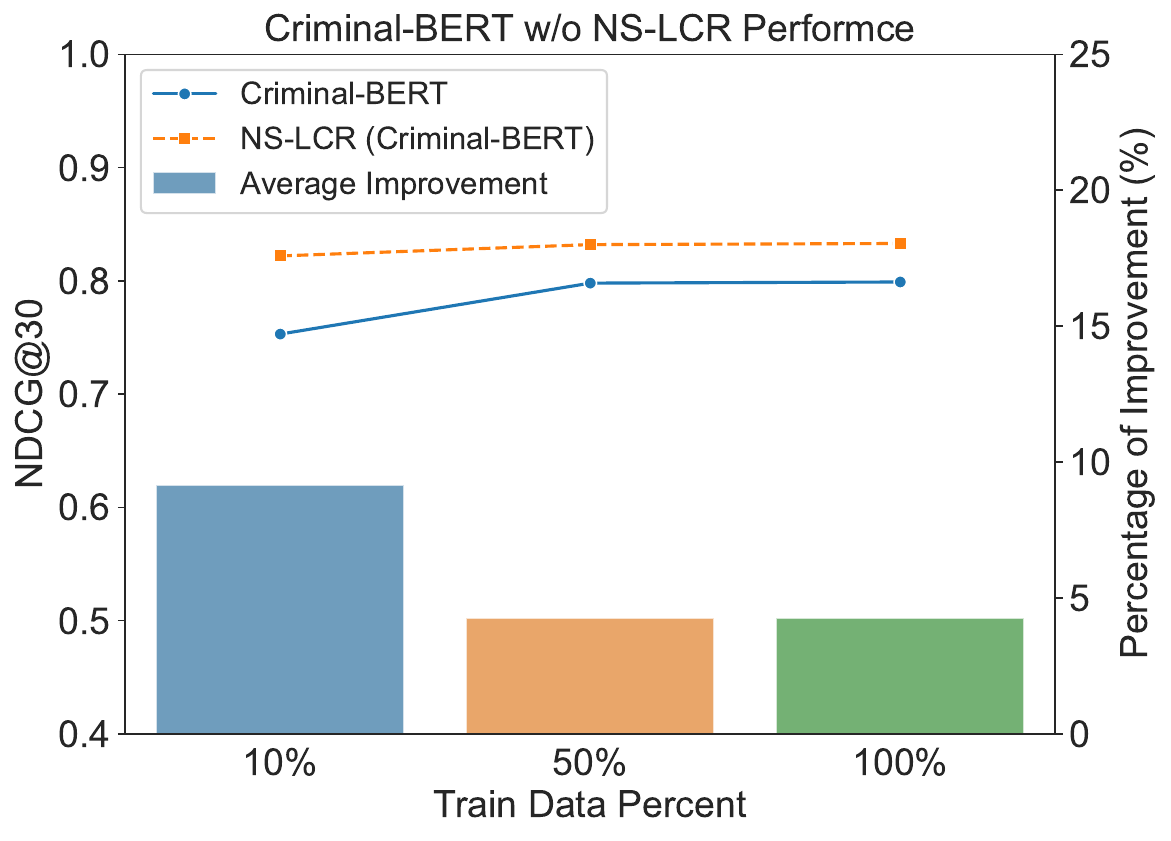}
  \end{minipage}
  \hfill
  \begin{minipage}[t]{0.49\linewidth}
    \includegraphics[width=\linewidth]{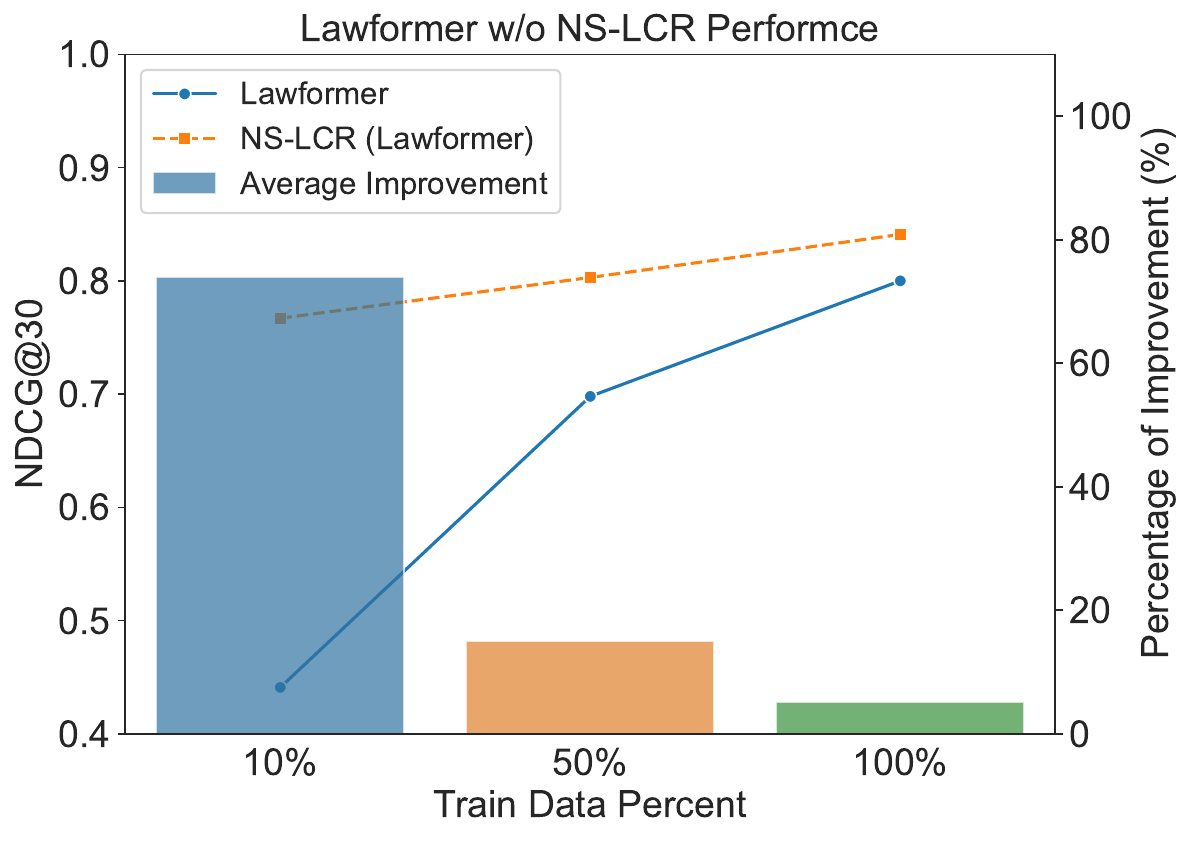}
  \end{minipage}
  
  \vspace{1em} 
  
  \begin{minipage}[t]{0.49\linewidth}
    \includegraphics[width=\linewidth]{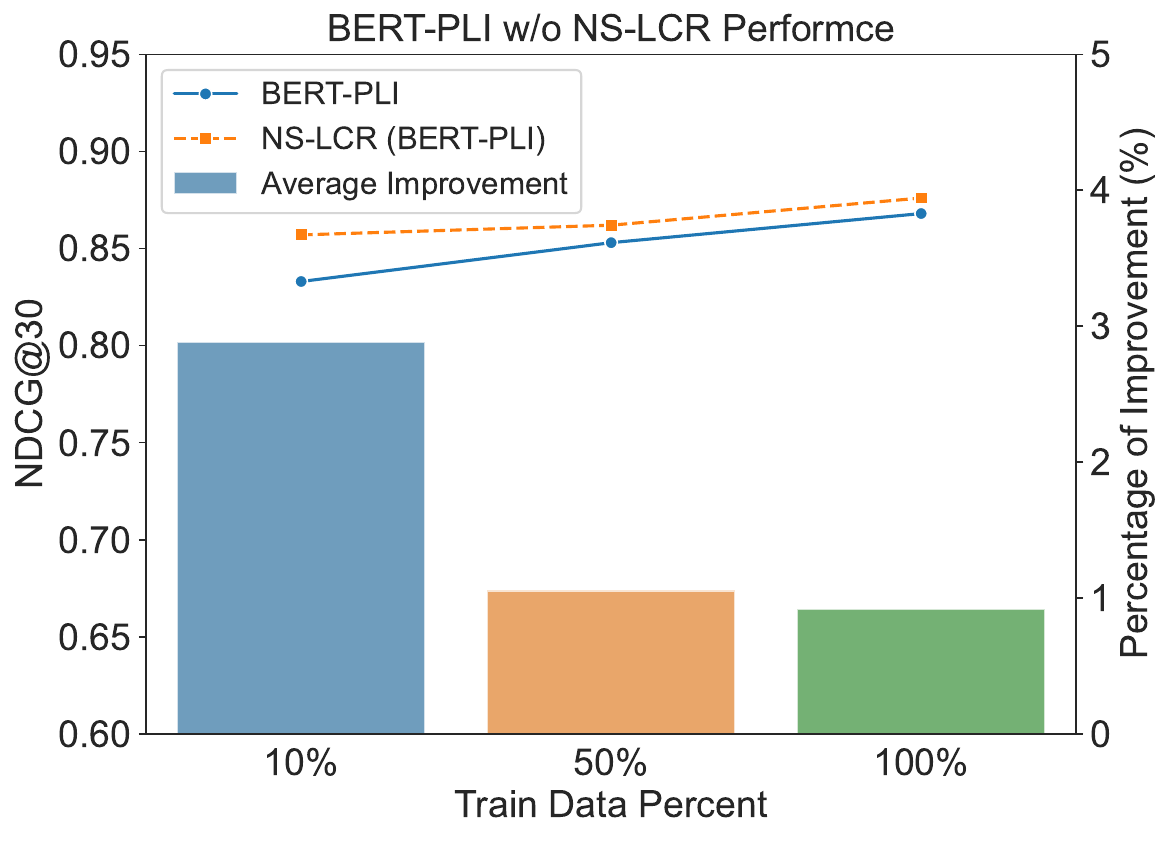}
  \end{minipage}
  \hfill
  \begin{minipage}[t]{0.49\linewidth}
    \includegraphics[width=\linewidth]{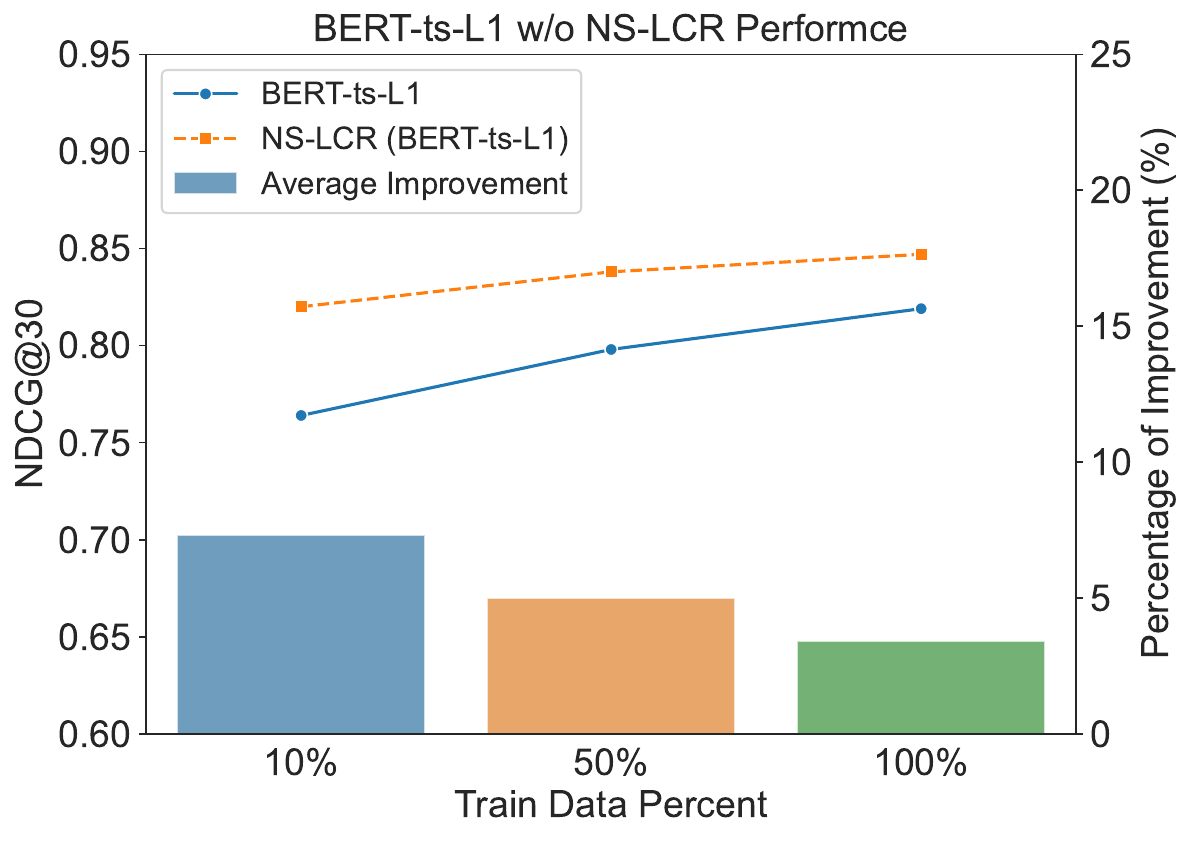}
  \end{minipage}

  \caption{Various base models' performance w/ and wo/ NS-LCR on LeCaRD is shown. The lines show the NDCG@30 scores and the bars show the improvement percentage due to NS-LCR.}
\label{fig:low-resource}
\end{figure}

In this section, we assess if NS-LCR can address the challenge of limited labeled data in LCR, given the high cost of expert annotation and diverse data distributions~\textsuperscript{\ref{fn:only_lecard}}. Through experiments with varied training data amounts, Figure~\ref{fig:low-resource} shows NS-LCR consistently improved performance across all base models at 10\%, 50\%, and 100\% LeCaRD training data proportions. The most notable gains occurred with less data, underscoring the advantage of our logic module in low-resource contexts. 
We also employed a hybrid approach that combines human evaluation and GPT-4~\cite{openai2023gpt4} to evaluate the explanations~\cite{bills2023language} directly. Specifically, we randomly sampled 50 queries from ELAM, retrieved the relevant candidate cases, and generated the explanations from IOT-Match, gpt-3.5-turbo~\footnote{The prompt template for generating the explanations is: \textit{``Please confirm whether the query and the candidate case below are relevant, and provide an explanation: [query];[candidate case]''}}, and NS-LCR. Then, we asked three annotators to judge the quality of explanations from IOT-Match, gpt-3.5-turbo, and NS-LCR based on the degree of alignment between the explanations, the original texts~(query and candidate case), and the law articles. At the same time, we used GPT-4 to score the quality of explanations from IOT-Match, gpt-3.5-turbo, and NS-LCR~\footnote{The prompt is: \textit{``Rate the explanations for IOT-Match, ChatGPT, and NS-LCR separately, with scores ranging from 0 to 10:''}.}.
As shown in~\autoref{tab:human_eval}, both the manual scores from the three annotators and the scores from GPT-4, NS-LCR, and gpt-3.5-turbo consistently achieved higher scores than IOT-Match. 
 Besides, gpt-3.5-turbo has not undergone legal task pre-training, and has a relatively limited understanding of legal knowledge~\cite{chalkidis2023chatgpt}, resulting in slightly lower performance compared to NS-LCR. In summary, the results verify the effectiveness of NS-LCR in generating high-quality explanations.

\begin{table}[t]
\small
  \centering
  \caption{Human and LLM evaluations of the explanation quality over 50 randomly sampled queries from ELAM (score range is from 0~(low) to 10~(high)~).}
  \begin{tabular}{l|c|c|c}
    \toprule
    & IOT-Match & gpt-3.5-turbo & NS-LCR \\
    \midrule
    Human & 7.74 & 8.75 & \textbf{8.95}\\
    \midrule
    GPT-4 & 7.86 & 8.71 & \textbf{8.86} \\
    \bottomrule
  \end{tabular}
  \label{tab:human_eval}
\end{table}

\subsection{Efficiency of NS-LCR}

\begin{table}[h]
    \centering
    \setlength{\abovecaptionskip}{0.3cm}
    \caption{Average online inference time per query, for four base models w/ or w/o NS-LCR.
    }
    \resizebox{0.95\linewidth}{!}{
    \begin{tabular}{l|c|c|c|c}
        \toprule
        Model&Criminal-BERT&BERT-ts-L1&Lawformer&Bert-PLI\\
        \midrule
         w/o NS-LCR & 1.5954 (s) &2.0107 (s) & 9.1730 (s) & 15.5668 (s)\\
         w/ NS-LCR& 3.6545 (s) & 4.0698 (s)& 11.2321 (s)& 17.6259 (s) \\
        \midrule 
         RelaCost & 129 \% & 102 \% & 22 \% & 13 \% \\
         \bottomrule
    \end{tabular}
    }
    \label{tab:time analysis}
\end{table}

We also analyze the efficiency of NS-LCR to demonstrate the additional time needed when conducting online inference. Specifically, we record the time required to process each query in the online inference stage, with different base models both with and without the NS-LCR module. Combined with ~\autoref{tab:time analysis} and ~\autoref{tab:main results}, it can be observed that the better the base model performs, the longer the time it consumes. Simultaneously, the relative time increase after adding NS-LCR is smaller. It can also be observed that although NS-LCR has a relatively large time increase on Criminal-BERT~\cite{zhong2019openclap} and BERT-ts-L1~\cite{shao2022understanding}, the total time is still far less than that of Lawformer and Bert-PLI. Moreover, NS-LCR achieves performance competitive with Lawformer~\cite{xiao2021lawformer} and Bert-PLI~\cite{shao2020bert}, thereby verifying its effectiveness.

\section{Conclusion}


This paper introduces the NS-LCR that neuro-symbolically combines case and law logic rules, providing faithful explainability. It aligns with different legal retrieval models. By updating LeCaRD and ELAM benchmarks with logic rules and introducing an explainability metric using LLMs, results show NS-LCR's excellent performance, reliable explanations, and improvement of base models in limited-resource scenarios.

\nocite{*}
\newpage
\section{Acknowledgements}
This work was funded by the National Key R\&D Program of China (2023YFA1008704), the National Natural Science Foundation of China (No.62376275, No.62377044), Beijing Key Laboratory of Big Data Management and Analysis Methods, Major Innovation \& Planning Inter-disciplinary Platform for the “Double-First Class” Initiative, funds for building world-class universities (disciplines) of Renmin University of China. Supported by the Outstanding Innovative Talents Cultivation Funded Programs 2024 of Renmin University of China.
\section{Bibliographical References}\label{sec:reference}

\bibliographystyle{lrec-coling2024-natbib}
\bibliography{lrec-coling2024-example}

\end{document}